\documentclass[aps,pra,twocolumn,showpacs]{revtex4}
\usepackage{graphicx,amsmath,amssymb,enumerate}
\begin{document}
\title{Discriminating Quantum Optical Beam Splitter Channels with Number-Diagonal-Signal States: Applications to Quantum Reading \\and Target Detection}
\author{Ranjith Nair\footnote{Email address: rnair@mit.edu}}
\affiliation{Research Laboratory of Electronics, Massachusetts Institute of Technology, Cambridge, MA 02139, USA}
\newcommand\mbf[1]{\mathbf{#1}}
\newcommand\ovl[1]{\overline{#1}}
\newcommand\udl[1]{\underline{#1}}
\newcommand\tld[1]{\tilde{#1}}
\newcommand {\ket}[1] {|{#1}\rangle}
\newcommand {\kets} [1] {|#1\rangle_S}
\newcommand {\keti} [1] {|#1\rangle_I}
\newcommand {\kete} [1] {|#1\rangle_E}
\newcommand {\bra}[1] {\langle{#1}|}
\newcommand {\bras}[1] {_S\langle{#1}|}
\newcommand {\brai}[1] {_I\langle{#1}|}
\newcommand {\brae}[1] {_E\langle{#1}|}
\newcommand{\braket}[2]{\langle{#1}|{#2}\rangle}
\newcommand {\nil}{\emptyset}
\newcommand {\cl}{\mathcal}
\newcommand {\tsf} [1]{\textsf{#1}}
\newcommand {\norm} [1] {\parallel #1 \parallel}
\newcommand {\al} {\alpha}
\newcommand {\e} {\epsilon}
\newcommand{\tr} {\tsf{tr}}
\newcommand {\beq} {\begin{equation}}
\newcommand {\eeq} {\end{equation}}
\newcommand {\bea} {\begin{align}}
\newcommand {\eea} {\end{align}}

\begin{abstract} We consider the problem of distinguishing with minimum probability of error two optical beam splitter channels with unequal complex-valued reflectivities using general quantum probe states entangled over $M$ signal \& $M'$ idler mode pairs of which the signal modes are bounced off the beam splitter while the idler modes are retained losslessly. We obtain a lower bound on the output state fidelity valid for \emph{any} pure input state. We define Number-Diagonal-Signal (NDS) States to be input states whose density operator in the signal modes is diagonal in the multimode number basis. For such input states, we derive series formulae for the optimal error probability, the output state fidelity, and the Chernoff-type upper bounds on the error probability.  For the special cases of quantum reading of a classical digital memory and target detection (for which the reflectivities are real-valued), we show that for a given input signal photon probability distribution, the fidelity is minimized by the NDS states with that distribution and that for a given average total signal energy $N_s$, the fidelity is minimized by any multimode Fock state with $N_s$ total signal photons. For reading of an ideal memory, it is shown that Fock state inputs minimize the Chernoff bound. For target detection under high loss conditions, a no-go result showing the lack of appreciable quantum advantage over coherent state transmitters is derived. A comparison of the error probability performance for quantum reading of number state and two-mode squeezed vacuum state (or EPR state) transmitters relative to coherent state transmitters is presented for various values of the reflectances. While the nonclassical states in general perform better than the coherent state, the quantitative performance gains differ depending on the values of the reflectances. The experimental outlook for realizing nonclassical gains from number state transmitters with current technology at moderate to high values of the reflectances is argued to be good.
\end{abstract}
\maketitle
\section{Introduction}\noindent
Consider, for $\tsf{b} \in \{0,1\},$ an optical beam splitter channel taking an input signal mode annihilation operator $\hat{a}_\tsf{in}$ to the output mode annihilation operator $\hat{a}_\tsf{out}$ in the manner of Fig.~1 under the mode transformation:-
\beq \label{beamsplitter}
\begin{pmatrix}
\hat{a}_\tsf{out} \\
\hat{e}_\tsf{out}
\end{pmatrix} =
\begin{pmatrix}
r_\tsf{b} e^{i\theta_{\tsf{b}}} & t_\tsf{b} \\
t_\tsf{b} e^{i\theta_\tsf{b}} & -r_\tsf{b}
\end{pmatrix}
\begin{pmatrix}
\hat{a}_\tsf{in} \\
\hat{e}_\tsf{in}
\end{pmatrix}.
\eeq
Here, $r_\tsf{b} = \sqrt{R_\tsf{b}}$ and $t_\tsf{b}= \sqrt{T_\tsf{b}}$ are real and non-negative field reflectivities with $R_\tsf{b}$ and $T_\tsf{b}$ the corresponding reflectances and transmittances with $R_\tsf{b} + T_\tsf{b}=1$. We will assume $R_0 \leq R_1$ throughout the paper. We have included, for later purposes, the input and output annihilation operators $\hat{e}_\tsf{in}$ and $\hat{e}_\tsf{out}$ of the other (environment) mode incident at the beam splitter, whose input is assumed to be in the vacuum state. Denoting the quantum channels induced on the signal mode by the above beam splitter transformation by $\cl{E_\tsf{b}}, \tsf{b} = 0/1$, and assuming that each of these channels has equal a priori probability, we consider the following general strategy to discriminate these two channels with minimum error probability illustrated in Fig.~1:-  A quantum state source S prepares a pure state $\ket{\psi}$  on the system consisting of $M$ `signal' optical modes $\{\hat{a}_\tsf{in}^m\}_{m=1}^M$ and $M'$ `idler' modes $\{\hat{b}_\tsf{in}^{m'}\}_{m'=1}^{M'}$, allowing for any entanglement across the signal modes and between the signal and idler modes. The $M$ signal modes are passed through $\cl{E}_\tsf{b}$ with the idler modes unchanged, giving rise, in general, to mixed density operators
\beq \label{rhob} \rho_\tsf{b} = \cl{E}_\tsf{b}^{\otimes^M} \otimes \tsf{id}^{\otimes^{M'}} (\ket{\psi}\bra{\psi})
\eeq
on the signal-idler Hilbert space depending on the value of $\tsf{b}$ -- here, $\tsf{id}$ denotes the identity channel on the idler modes. This joint state is now measured at the detector D using the Helstrom quantum measurement that yields minimum error probability in distinguishing the two states \cite{helstrom76}.
\begin{figure}
\includegraphics[trim= 20mm 24mm 20mm 30mm, clip=true, width=0.5\textwidth]{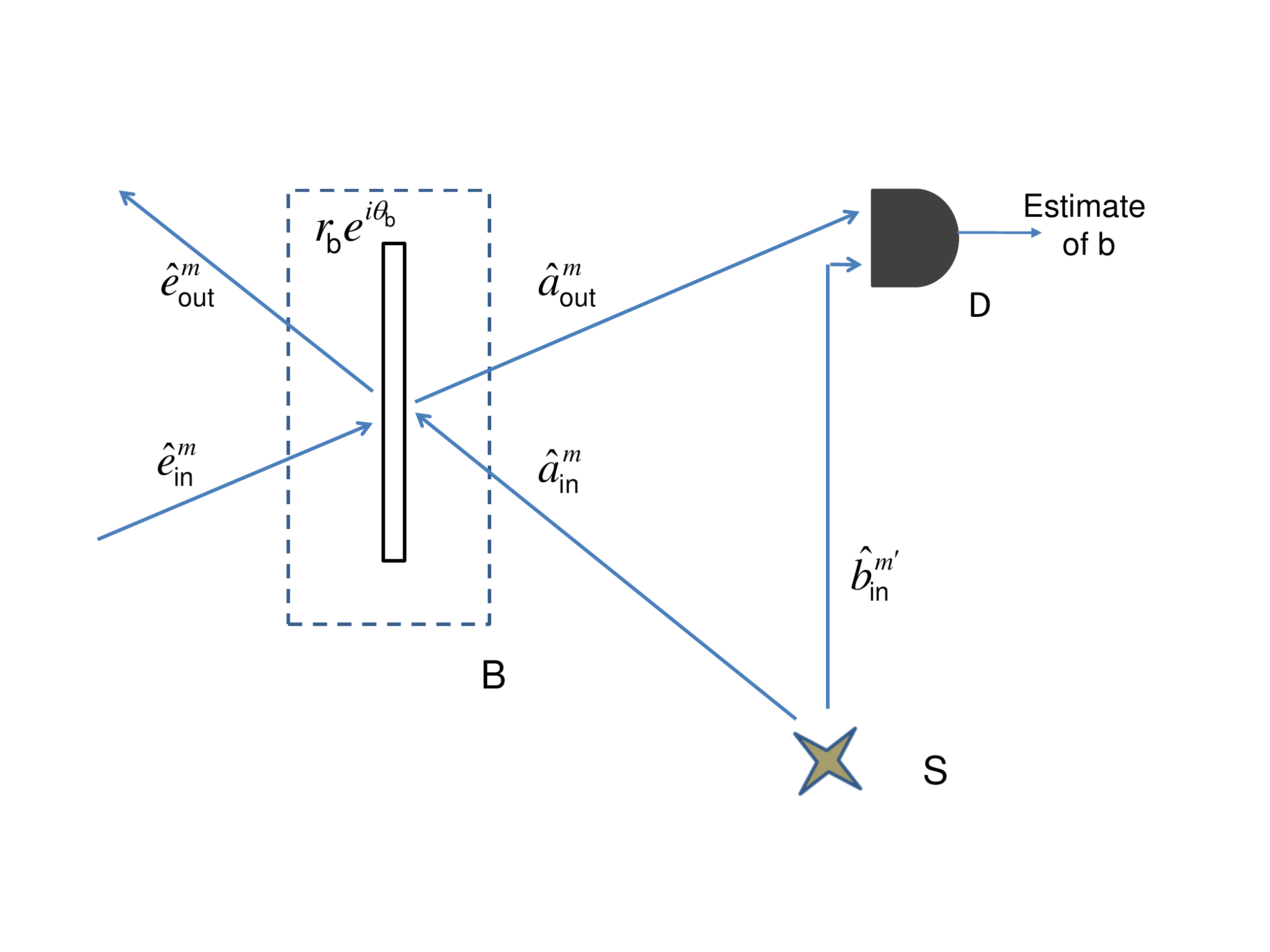}
\caption{Schematic of setup for determining which of two beam splitter channels (indexed by $\tsf{b}$) of the form of \eqref{beamsplitter} is present within a black box B. A quantum state source S produces a pure state of $M$ signal modes $\{\hat{a}_\tsf{in}^m\}_{m=1}^M$ and $M'$ idler modes $\{\hat{b}_\tsf{in}^{m'}\}_{m'=1}^{M'}$ of which one signal-idler pair is shown. Each signal mode reflects off the beam splitter while the idler mode is unaffected. The environment modes $\{\hat{e}_\tsf{in}^m\}_{m=1}^M$ coupling to the beam splitter are all in the vacuum state. The detection module D performs the minimum error probability (Helstrom) measurement for distinguishing the two possible $(M+M')$-mode quantum states corresponding to $\tsf{b} = 0/1$.}
\end{figure}
In view of the fact that loss is ubiquitous in quantum state transmission, processing and detection, and because of the well-known sensitivity of nonclassical states to loss, we will in general assume that $r_\tsf{b} < 1$. On the other hand, at room temperature and at optical wavelengths $\sim 1 \hspace{1mm}\mu$m, the average number of thermal noise photons per space-time mode is ideally around $10^{-21}$, allowing us realistically to neglect it, although it is known that interesting nonclassical gains are obtained in high-loss high-noise conditions such as those in `quantum illumination' \cite{Tan08}.

Many interesting problems are encompassed by the above model. The case of $r_0=0$ (and possibly  $r_1 \ll 1$) corresponds to a target detection scenario (identical to quantum illumination \cite{Tan08} but for the absence of thermal noise) in which $\tsf{b}=0,1$ represents the absence or presence of a reflecting target of effective field reflectivity $r_1$. The more general case of non-zero $r_0$ but $\Delta\equiv \theta_1 - \theta_0 =0$ corresponds to the recently proposed quantum reading of a classical digital memory \cite{Pirandola11}. The case of $r_0=r_1 <1, \Delta \neq 0$ models the lossy discrimination of channels differing only in phase shift (or path length), and is related to the problem of estimation of a continuous optical phase parameter.

Given this background, the following question may now be posed: Among all $\ket{\psi}$'s with average total energy $N_s$ in the signal modes, which state \emph{minimizes} the average error probability in distinguishing the channels, and how much improvement over a \emph{classical state} of the same energy is attainable? A multimode classical state is a state with a non-negative P-representation \cite{sudarshan63,glauber63}. Thus $\rho$ given by
\begin{align} \label{classicalstate}
\rho= \int   P(\boldsymbol{\alpha}) \ket{\boldsymbol{\alpha}} \bra{\boldsymbol{\alpha}} d^2\boldsymbol{\alpha}
\end{align}
is classical if $P(\boldsymbol{\alpha})\geq 0$ for $\ket{\boldsymbol{\alpha}}$ an $M$-mode coherent state, so that $\rho$ is a probabilistic mixture of product coherent states. Comparison of the performance of a proposed input state to this class of states is important since, as their definition implies, the latter are readily prepared by modulating laser fields with classical random numbers. In particular, we will compare performance with the pure coherent state $\ket{\sqrt{N_s}}\otimes\ket{0}^{\otimes^{M-1}}$ \cite{note1}.

A related version of the general problem of discriminating two beam splitter channels has been addressed recently in \cite{Bisio11}. The scenario of \cite{Bisio11} differs from ours in that it is assumed, referring to Fig.~1, that the \emph{combined} state of $\hat{e}_{\tsf{in}}$ and $\hat{a}_{\tsf{in}}$ may be chosen subject to a total energy constraint and that \emph{both} the beam splitter output modes $\hat{a}_{\tsf{out}}$ and $\hat{e}_{\tsf{out}}$ are available for a Helstrom detector to measure, in addition to the idler mode $\hat{b}_{\tsf{in}}$.  As such, the problem of \cite{Bisio11} is one of discrimating two unitary transformations as opposed to the non-unitary channel discrimination problem considered here. Nevertheless, it is clear that, for $M=1$, the minimum error probability attainable at given $N_s$ in the setting of \cite{Bisio11} is not greater than that for our more restrictive setting. On the other hand, the case of $M>1$ was not considered in \cite{Bisio11}. We note that the problem in the form addressed here is of more relevance to long-range scenarios where the black box B of Fig.~1, and therefore, the input and output environment modes, are not directly accessible to the user who controls the state source S and the detector D. It also models quantum reading scenarios where the memory itself (e.g., a CD or DVD) need not be modified in a major way. When such modifications to the memory are contemplated, the scenario of \cite{Bisio11} becomes interesting.

This paper provides several partial answers to the broad question posed above.  In Section II, we derive a lower bound on the Jozsa-Uhlmann fidelity \cite{Uhlmann76,Jozsa94} between $\rho_0$ and $\rho_1$
\beq \label{fidelity}
 \cl{F} = \mathrm{Tr} ^2 \sqrt{\sqrt{\rho_0}\rho_1\sqrt{\rho_0}}
\eeq
that is valid for \emph{any} input state $\ket{\psi}$ \cite{noteondef}. This further yields a lower bound on the minimum error probability of distinguishing the output states. In Section III, we specialize to input states whose reduced density operator on the signal modes is a mixture of multimode number states -- such states will be referred to as Number-Diagonal-Signal (NDS) States in this paper. It was shown recently in \cite{NairYen11} that NDS states are optimum input states according to many possible performance criteria for a large class of image sensing problems of which the minimum error probability discrimination between two beam splitter channels that is the subject of the present paper is a special case. We mention that, in the contexts of communication and key distribution between two users, a class of states related to the NDS states was introduced in \cite{UsenkoParis07} and referred to as Photon-Number Entangled States (PNES). A PNES is a pure state of a single signal-idler mode pair that is diagonal in the number state basis of both the signal and idler modes. Since the states we consider do not have to be number-state-diagonal in the idler modes, we believe the term ``Number-Diagonal-Signal" state is more appropriate. However, it is the case that a given single-mode NDS state can be related to a corresponding PNES by a unitary transformation on the idler modes without changing the error performance.

The NDS states include, but are not limited to, number states, the two-mode squeezed vacuum (or EPR) states used in \cite{Tan08,Pirandola11}, the NOON states \cite{Dowling08}, the related $|m::m'\rangle$ states of \cite{Huver08}, the pair coherent states (PCS) \cite{Agarwal86}, and several other states studied in the context of phase estimation \cite{Lee09}.
For this class of states, we show that $\rho_0$ and $\rho_1$ of \eqref{rhob} are easily diagonalized. The minimum possible error probability of discriminating any two states (with equal a priori probabilities) is given by the Helstrom formula \cite{helstrom76}
\beq \label{helstrom}
\ovl{P}_e =  \frac{1}{2} - \frac{1}{4} \norm{\rho_0 - \rho_1}_1,
\eeq
where $\norm{A}_1$ is the trace-norm which equals, for self-adjoint $A$, the sum of the absolute values of the eigenvalues of $A$. For non-commuting mixed states, $\ovl{P}_e$ is notoriously hard to calculate. For input NDS states, however, we show in Section III that the eigenvectors of $\rho_0$ and $\rho_1$ have a mutual inner product structure that permits the calculation of $\ovl{P}_e$ as a, in general infinite, series. The fidelity \eqref{fidelity} also yields the following upper and lower bounds on  $\ovl{P}_e$  \cite{Fuchs99} (the first inequality follows from a binomial expansion of its RHS):-
\beq \label{fidelityboundsonhelstrom}
\frac{1}{4}\cl{F} \leq \frac{1}{2}(1- \sqrt{1-\cl{F}}) \leq \ovl{P}_e \leq \frac{1}{2} \sqrt{\cl{F}}.
\eeq
Also of interest are the ``Chernoff-type'' upper bounds:-
\beq \label{chernoff}
\ovl{P}_e \leq \frac{1}{2} Q(s), \hspace{2mm}s \in [0,1],
\eeq
where $Q(s) = \mathrm{Tr}[\rho_0^s \rho_1^{1-s}]$ and the \textit{Chernoff bound}  $Q$ \cite{Audenaert06} is the best such bound:-
\begin{align}
\ovl{P}_e \leq \frac{1}{2} Q,
\end{align}
where
\begin{align}
Q = \min_{s \in [0,1]} Q(s).
\end{align}
The \textit{Bhattacharyya bound} is \eqref{chernoff} with $s=1/2$. This terminology was introduced in \cite{Pirandola08} and the bound was applied for the first time in \cite{Tan08}. We show here that, for NDS input states, $\cl{F}$ and $Q(s)$ are also calculable as infinite series in general, and have closed-form expressions in some cases.

In Section IV, we apply these results to quantum reading and target detection, i.e., to the cases where $\Delta=0$.  The input state consisting of  $M$ signal and idler mode pairs from the output of a parametric downconversion process was shown in \cite{Pirandola11} to yield, in some regions of $M$ and $N_s$, surprisingly better error probability in quantum reading than that obtainable from any classical state of the same energy \cite{caveat}. In this paper, we first show in Section IV.A that, for any given signal photon number distribution, the general fidelity lower bound derived in Section I is attained by NDS states. Further, the NDS states minimizing the fidelity for a given $N_s$ and $M$ are the multimode Fock states of total photon number $N_s$. In Section IV.A.1, we lower bound the error probability for quantum reading and target detection via the fidelity for general input states. In Section IV.A.2, for the case of $R_1=1$, we show that the Fock states give the lowest Chernoff bound among all pure input states of given energy. In Section IV.A.3, we state a fidelity-based error probability lower bound for target detection with a general input state. In the limit $R_1 \ll 1$ of large loss, we derive a no-go result that rules out appreciable quantum advantage over coherent states.

In Sections IV.B-D, we characterize in detail the error probability performance of coherent states, number states and the EPR states. We show that the performance of all the multimode Fock states of total photon number $N_s$ are identical, a fact that helps practical implementation. For EPR states, we show that the application of the results of Section III yields the same analytical results as the methods used in \cite{Pirandola11}. In Section IV.E.1, we compare the quantitative performance of coherent, number, and EPR states for some typical reflectance values and demonstrate nonclassical gains. In Section IV.E.2, we consider the technological feasibility of achieving the nonclassical gains in quantum reading.

\section{Lower Bound on Output State Fidelity}  Consider an arbitrary $(M'+M)$-mode state $\ket{\psi}$ in the photon number representation:-
\beq
\ket{\psi} = \sum_{\mbf{m}, \mbf{n}} c_{\mbf{m}, \mbf{n}} \ket{\mbf{m}}_I \ket{\mbf{n}}_S,
\eeq
where $\keti{\mbf{m}} = \ket{m_1}_{I_1} \otimes \cdots \otimes \ket{m_{M'}}_{I_{M'}} $ and $\kets{\mbf{n}} =\ket{n_1}_{S_1} \otimes \cdots \otimes \ket{n_M}_{S_M} $ are respectively the photon number states of the $M/M'$-mode signal/idler. After augmenting the above state with the vacuum $\kete{\mbf{0}}$ of the $M$ environment modes $\{\hat{e}_\tsf{in}\}_{m=1}^M$ of \eqref{beamsplitter}, we may write down the Schr\"{o}dinger-picture evolution corresponding to the Heisenberg-picture evolution \eqref{beamsplitter} of the joint system to one of two pure states as:-
\beq \label{purification}
\ket{\psi^{(\tsf{b})}} =  \sum_{\mbf{m}, \mbf{n}} c_{\mbf{m}, \mbf{n}} \sum_{\mbf{k} \leq \mbf{n}} A_{\mbf{n};\mbf{k}}^{(\tsf{b})} \hspace{1mm}\ket{\mbf{m}}_I \ket{\mbf{n}-\mbf{k}}_S \ket{\mbf{k}}_E,
\eeq
where the amplitude $A_{\mbf{n};\mbf{k}}^{(\tsf{b})}$ is given by:-
\beq
A_{\mbf{n};\mbf{k}}^{(\tsf{b})} =  \prod_{m=1}^{M} \left[ \sqrt{\binom {n_m}{k_m}} \hspace{1mm} e^{i n_m \theta_\tsf{b}} \hspace{1mm} r_\tsf{b}^{n_m - k_m} t_\tsf{b}^{k_m}\right].
\eeq
In \eqref{purification}, $\mbf{k} \leq \mbf{n}$ is to be understood as component-wise inequality. We then have $\rho_{\tsf{b}} = \mathrm{Tr}_E [\ket{\psi^{(\tsf{b})}}\bra{\psi^{(\tsf{b})}}]$ so that the $\ket{\psi^{(\tsf{b})}}$ are purifications \cite{nielsen00} of the respective $\rho_\tsf{b}$. From \eqref{purification}, it follows that the overlap $O=|\braket {\psi^{(0)}} {\psi^{(1)}}|^2$ equals
\begin{align}
&O=\\ & \left|\sum_{\mbf{m}, \mbf{n}} |c_{\mbf{m}, \mbf{n}}|^2  \sum_{\mbf{k} \leq \mbf{n}} \prod_{m=1}^{M} \left[ \binom {n_m}{k_m} \hspace{1mm} e^{i n_m \Delta} \hspace{1mm} (r_0 r_1)^{n_m - k_m} (t_0 t_1)^{k_m}\right]\right|^2 \\
 =& \left|\sum_{\mbf{n}} p_{\mbf{n}} \sum_{\mbf{k} \leq \mbf{n}} \prod_{m=1}^{M} \left[ \binom {n_m}{k_m} \hspace{1mm} e^{i n_m \Delta} \hspace{1mm} (r_0 r_1)^{n_m - k_m} (t_0 t_1)^{k_m}\right]\right|^2 \\
 =& \Bigg|\sum_{\mbf{n}} p_{\mbf{n}} \prod_{m=1}^{M} \left[ e^{i n_m \Delta} (r_0r_1 + t_0 t_1)^{n_m} \right]\Bigg|^2 \label{Osemifinal}\\
 =& \Big|\sum_{n=0}^{\infty} p_{n} e^{i n \Delta} (r_0r_1 + t_0 t_1)^{n}\Big|^2, \label{Ofinal}
\end{align}
where $p_{\mbf{n}} = \sum_{\mbf{m}} |c_{\mbf{m}, \mbf{n}}|^2$ is the multimode photon  probability distribution in the signal modes. In going from \eqref{Osemifinal} to \eqref{Ofinal}, we have re-ordered the sum of \eqref{Osemifinal} over the single index $n = \sum_{m=1}^{M} n_m$ with $p_n$ being the probability distribution of the \emph{total} signal photon number:-
\begin{align} \label{pndef}
p_n = \sum_{\mbf{n}: \sum_{m=1}^M n_m = n} p_{\mbf{n}},
\end{align}
Uhlmann's theorem \cite{Jozsa94,Uhlmann76,nielsen00} states that the fidelity $\cl{F}$ is the \emph{maximum} overlap over all purifications of $\rho_0$ and $\rho_1$, so that we have the lower bound
\beq \label{fidelitylowerbound}
O \leq \cl{F}
\eeq
for the $O$ of \eqref{Ofinal}. For any proposed input state $\ket{\psi}$, the overlap $O(\ket{\psi})$ can be calculated via \eqref{Ofinal} in terms of the signal photon probability distribution of $\ket{\psi}$. Application of the inequalities \eqref{fidelitylowerbound} and \eqref{fidelityboundsonhelstrom} yields the following lower bound on $\overline{P}_e$:-
\begin{align}
\frac{1}{2}\left(1- \sqrt{1-O(\ket{\psi})}\right) \leq \overline{P}_e[\ket{\psi}].
\end{align}

The cases of quantum reading and target detection for which $\Delta= 0$ are further developed in Section IV.A.

\section{Number-Diagonal-Signal (NDS) States: Minimum Error Probability, Output State Fidelity, and Chernoff-Type Bounds} \noindent  We define Number-Diagonal-Signal (NDS) States to be states whose reduced density operator in the $M$ signal modes is diagonal in the product number (Fock) state basis. A pure NDS state then has the representation
\beq
\ket{\psi}= \sum_{\mbf{n}} c_{\mbf{n}} \keti{\phi_\mbf{n}}\kets{\mbf{n}},
\eeq
where $\keti{\phi_\mbf{n}}$ is \emph{any} orthonormal set of states on the idler modes -- the above equation is essentially a Schmidt decomposition \cite{nielsen00} of $\ket{\psi}$.
This is a wide class of states, many of which have been intensively studied in quantum optics and have interesting applications in quantum information and metrology \cite{Tan08,Pirandola11,UsenkoParis07,Dowling08,Huver08,Agarwal86,Lee09}.

We now show that, for NDS inputs, $\rho_0$ and $\rho_1$ have a form helpful for calculations.  After propagation through the beam splitter channel we obtain, as in \eqref{purification}, the purifications
\begin{align}
\ket{\psi^{(\tsf{b})}} =&  \sum_{\mbf{n}} c_{\mbf{n}} \sum_{\mbf{k} \leq \mbf{n}} A_{\mbf{n};\mbf{k}}^{(\tsf{b})} \hspace{1mm}\ket{\phi_\mbf{n}}_I \ket{\mbf{n}-\mbf{k}}_S \ket{\mbf{k}}_E \\
=& \sum_{\mbf{k}} \Big( \sum_{\mbf{n}: \mbf{n} \geq \mbf{k}} c_{\mbf{n}} A_{\mbf{n};\mbf{k}}^{(\tsf{b})} \hspace{1mm}\ket{\phi_\mbf{n}}_I \ket{\mbf{n}-\mbf{k}}_S\Big)\ket{\mbf{k}}_E \\
\equiv & \sum_{\mbf{k}} \ket{\psi_{\mbf{k}}^{(\tsf{b})}} \ket{\mbf{k}}_E,
\end{align}
where
\begin{align} \label{psikb}
\ket{\psi_{\mbf{k}}^{(\tsf{b})}} = \sum_{\mbf{n}: \mbf{n} \geq \mbf{k}} c_{\mbf{n}} A_{\mbf{n};\mbf{k}}^{(\tsf{b})} \hspace{1mm}\ket{\phi_\mbf{n}}_I \ket{\mbf{n}-\mbf{k}}_S
\end{align}
are un-normalized joint signal-idler states.
Since the $\{\ket{\phi_{\mbf{n}}}_I\}$ are an orthonormal set by the NDS state assumption, we see that the orthogonality relations
\begin{align} \label{selfip}
\braket{\psi_{\mbf{k}}^{(\tsf{b})}}{\psi_{\mbf{k'}}^{(\tsf{b})}} = p_{\mbf{k}}^{(\tsf{b})} \delta_{\mbf{k},\mbf{k'}},
\end{align}
hold because the $\ket{\psi_{\mbf{k}}^{(\tsf{b})}}$ and $\ket{\psi_{\mbf{k'}}^{(\tsf{b})}}$ of \eqref{psikb} are termwise orthogonal when $\mbf{k} \neq \mbf{k'}$.  Here $p_{\mbf{k}}^{(\tsf{b})}$ is the probability, conditioned on $\tsf{b}$, of finding $\mbf{k}$ photons in the output environment modes and is given by
\begin{align} \label{pkb}
p_{\mbf{k}}^{(\tsf{b})} =& \sum_{\mbf{n}: \mbf{n} \geq \mbf{k}} |c_{\mbf{n}} A_{\mbf{n};\mbf{k}}^{(\tsf{b})}|^2 \equiv \sum_{\mbf{n}: \mbf{n} \geq \mbf{k}} p_{\mbf{n}} |A_{\mbf{n};\mbf{k}}^{(\tsf{b})}|^2 \\
=&\sum_{\mbf{n}: \mbf{n} \geq \mbf{k}} p_{\mbf{n}}\prod_{m=1}^{M} \left[ \binom {n_m}{k_m} \hspace{1mm}  \hspace{1mm} R_\tsf{b}^{(n_m - k_m)} T_\tsf{b}^{k_m}\right].
\end{align}
Again by the NDS state assumption, we have
\begin{align} \label{mutualip}
\braket{\psi_{\mbf{k}}^{(\tsf{0})}}{\psi_{\mbf{k'}}^{(\tsf{1})}} =  I_{\mbf{k}} \cdot \delta_{\mbf{k},\mbf{k'}},
\end{align}
for
\begin{align} \label{Ik}
I_{\mbf{k}} =& \sum_{\mbf{n}: \mbf{n} \geq \mbf{k}} |c_{\mbf{n}}|^2 (A_{\mbf{n};\mbf{k}}^{(\tsf{0})})^* A_{\mbf{n};\mbf{k}}^{(\tsf{1})} \\
=& \sum_{\mbf{n}: \mbf{n} \geq \mbf{k}} p_{\mbf{n}}\prod_{m=1}^{M} \left[ \binom {n_m}{k_m} \hspace{1mm} e^{i n_m \Delta} \hspace{1mm} (r_0 r_1)^{n_m - k_m} (t_0 t_1)^{k_m}\right]. \label{Ik2}
\end{align}
The orthogonality relations \eqref{selfip} and \eqref{mutualip} are key to the mathematical tractability of the optimal error probability problem for NDS input states.

\subsection{Optimal (Helstrom) Detection: Error Probability and Measurement Operators} \noindent Tracing over the environment modes in (24), we obtain  $\rho_0$ and $\rho_1$, which are already in diagonal form by virtue of \eqref{selfip}:-
\begin{align}
\rho_{\tsf{b}} = \sum_{\mbf{k}} \ket{\psi_{\mbf{k}}^{(\tsf{b})}}\bra{\psi_{\mbf{k}}^{(\tsf{b})}}.
\end{align}
For $\cl{H}_\mbf{k} = \textrm{span}\{\psi_{\mbf{k}}^{(\tsf{0})},\psi_{\mbf{k}}^{(\tsf{1})}\}$, we have via \eqref{selfip} and \eqref{mutualip} that the 2-dimensional spaces $\{\cl{H}_\mbf{k}\}$ are mutually orthogonal. The joint signal-idler Hilbert space $\cl{H}$ may then be expressed as an orthogonal direct sum of the $\{\cl{H}_\mbf{k}\}$ with an additional component $\cl{H}_\perp$ orthogonal to all the $\cl{H}_\mbf{k}$:-
\begin{align} \label{hilbertspacedecomp}
 \cl{H} = \bigoplus_{\mbf{k}} \cl{H}_\mbf{k} \oplus \cl{H}_\perp.
\end{align}
For the purpose of distinguishing $\rho_0$ and $\rho_1$, we may restrict the domain of definition of the difference density operator $\Delta\rho \equiv \rho_0 - \rho_1$  appearing in the Helstrom formula \eqref{helstrom} to $\bigoplus_\mbf{k} \cl{H}_\mbf{k}$ since the support of both $\rho_0$ and $\rho_1$ is orthogonal to $\cl{H}_\perp$. Corresponding to \eqref{hilbertspacedecomp}, $\Delta\rho$ may be decomposed into a direct sum of operators on $\cl{H}_\mbf{k}$
\begin{align}
\Delta \rho = \bigoplus_{\mbf{k}} \Delta \rho\big| _{\mbf{k}}, \hspace{2mm}\Delta \rho\big| _{\mbf{k}} \in \cl{L}(\cl{H}_\mbf{k}),
\end{align}
where
\begin{align}
\Delta \rho\big| _{\mbf{k}}= \ket{\psi_{\mbf{k}}^{(0)}}\bra{\psi_{\mbf{k}}^{(0)}} - \ket{\psi_{\mbf{k}}^{(1)}}\bra{\psi_{\mbf{k}}^{(1)}}.
\end{align}
Performing a Gram-Schmidt orthonormalization on each $\cl{H}_\mbf{k} = \textrm{span}\{\psi_{\mbf{k}}^{(\tsf{0})},\psi_{\mbf{k}}^{(\tsf{1})}\}$ with $\psi_{\mbf{k}}^{(\tsf{0})}/ \norm{\psi_{\mbf{k}}^{(\tsf{0})}}$ as the first basis vector, we may write the $2 \times 2$ matrix of $\Delta\rho\big|_\mbf{k}$ in the Gram-Schmidt basis as \cite{note2}:-
\begin{align} \label{Deltarhok}
\Delta \rho\big| _{\mbf{k}}= \begin{pmatrix} p_\mbf{k}^{(0)} - {|I_\mbf{k}|^2}/{p_\mbf{k}^{(0)}} & -\frac{I_\mbf{k}}{p_\mbf{k}^{(0)}}\sqrt{p_\mbf{k}^{(0)}p_\mbf{k}^{(1)}-|I_\mbf{k}|^2}  \\
-\frac{I^*_\mbf{k}}{p_\mbf{k}^{(0)}}\sqrt{p_\mbf{k}^{(0)}p_\mbf{k}^{(1)}-|I_\mbf{k}|^2} & -p_\mbf{k}^{(1)} + |I_\mbf{k}|^2/p_\mbf{k}^{(0)}.
\end{pmatrix}
\end{align}
On calculating the eigenvalues and trace norm of the above matrix, the minimum error probability follows as:-
\begin{align}
\overline{P}_e =& \frac{1}{2} - \frac{1}{4}  \Big\|\bigoplus_{\mbf{k}}\left(\Delta\rho\big|_{\mbf{k}}\right)\Big\|_1 \\
=&\frac{1}{2} - \frac{1}{4} \sum_{\mbf{k}} \Big\|\Delta\rho\big|_{\mbf{k}}\Big\|_1 \\ \label{Pebarnds}
=&  \frac{1}{2} - \frac{1}{4} \sum_{\mbf{k}} \Big[ \big(p_\mbf{k}^{(0)}+p_\mbf{k}^{(1)}\big)^2 - 4|I_{\mbf{k}}|^2\Big]^{1/2}.
\end{align}
Although the above sum may be hard to evaluate analytically, numerical computation to any desired accuracy is always possible.

The discussion above also yields the abstract mathematical description of the measurement operators for optimally discriminating $\rho_0$ and $\rho_1$. The optimal measurement consists of two orthogonal projection operators $\Pi_0$ and $\Pi_1$  augmented with an additional projection operator $\Pi_\perp$ onto $\cl{H}_{\perp}$ to make a complete projective measurement with $\Pi_0 + \Pi_1 + \Pi_\perp = I_{IS},$ the identity on $\cl{H}$. Note that the state never projects onto $\cl{H}_\perp$. $\Pi_0$ and $\Pi_1$ are given in the usual way \cite{helstrom76} by
\begin{align} \label{Pi0}
\Pi_0 = \sum_{\mbf{k}} \Pi_{+} \{\Delta \rho \big|_{\mbf{k}}\}
\end{align}
and
\begin{align} \label{Pi1}
\Pi_1 = \sum_{\mbf{k}} \Pi_{-} \{\Delta \rho \big|_{\mbf{k}}\},
\end{align}
where $\Pi_{+/-} \{\Delta \rho \big|_{\mbf{k}}\}$ is the projection operator onto the one-dimensional eigenspace corresponding to the positive/negative eigenvalue of $\Delta \rho \big|_{\mbf{k}}$. This space is the span of the eigenvector of $\Delta \rho \big|_{\mbf{k}}$ of \eqref{Deltarhok} with positive/negative eigenvalue. In somewhat more physical language, we may view the optimum measurement as a two-stage measurement. In the first stage, we perform a Quantum Non-Demolition (QND) measurement, i.e., a measurement that projects the state on to one of several orthogonal Hilbert spaces without destroying it. In our case, the received state is projected into one of the orthogonal spaces $\cl{H}_\mbf{k}$. Depending on the value of $\mbf{k}$, we further make on $\cl{H}_{\mbf{k}}$ the binary Helstrom measurement corresponding to the optimum discrimination of $\ket{\psi^{(0)}_\mbf{k}}$ and $\ket{\psi^{(1)}_\mbf{k}}$ with conditional probabilities $p_\mbf{k}^{(0)}/(p_\mbf{k}^{(0)}+p_\mbf{k}^{(1)})$ and $p_\mbf{k}^{(1)}/(p_\mbf{k}^{(0)}+p_\mbf{k}^{(1)})$ respectively. From this description, we may expect that this measurement is hard to realize in the laboratory since the eigenvectors of $\Delta \rho \big|_{\mbf{k}}$ are $(M'+M)$-mode entangled states in general.

The case of unequal a priori probabilities, say $\pi_0$ and $\pi_1$ for $\rho_0$ and $\rho_1$, may be handled similarly. $\overline{P}_e$ is now given by
\begin{align} \label{unequalpriors}
\overline{P}_e = \frac{1}{2} - \frac{1}{2}|| \pi_0 \rho_0 - \pi_1 \rho_1||_1.
\end{align}
The operator $\pi_0 \rho_0 - \pi_1 \rho_1$ also has an orthogonal direct sum decomposition over the $\{\cl{H}_{\mbf{k}}\}$ (which are unchanged). However, the matrices for $\Delta \rho \big|_{\mbf{k}}$ are different from \eqref{Deltarhok} and include the prior probabilities. Diagonalizing these matrices yields the optimal error probability and the measurement operators.

\subsection{Fidelity and Chernoff-type Bounds} \noindent
The optimum error probability $\overline{P}_e$ is usefully bounded by the fidelity and Chernoff-type bounds \eqref{fidelityboundsonhelstrom} and \eqref{chernoff}. For NDS states, it is straightforward to show, using \eqref{selfip} and \eqref{mutualip} in the definition \eqref{fidelity}, that the fidelity is given by
\begin{align} \label{ndsfidelity}
\cl{F} =& \left(\sum_{\mbf{k}} |I_{\mbf{k}}|\right)^2 \\ \label{ndsfidelity2}
=& \left(\sum_{\mbf{k}} \left| \sum_{\mbf{n}: \mbf{n} \geq \mbf{k}} p_{\mbf{n}}\prod_{m=1}^{M} \binom {n_m}{k_m} \hspace{1mm} e^{i n_m \Delta} \hspace{1mm} (r_0 r_1)^{n_m - k_m} (t_0 t_1)^{k_m}\right| \right)^2
\end{align}
where we have used (\ref{Ik2}). The presence of the absolute value sign prevents simplification of the above expression without further assumptions, but we note that, for $\Delta \neq 0$, \eqref{ndsfidelity2} is greater and hence more informative, than the general lower bound (15).

The Chernoff-type quantities may be likewise computed to equal
\begin{align} \label{ndschernoffquantity}
Q(s) = \sum_{\mbf{k}} [p_\mbf{k}^{(0)}]^{s-1}[p_\mbf{k}^{(1)}]^{-s}|I_{\mbf{k}}|^2,
\end{align}
for $s \in [\hspace{0.5mm}0,1]$. The `singularity' in the terms for which $p_\mbf{k}^{(\tsf{b})}=0$ is only apparent as $I_{\mbf{k}}$ is also zero for those terms. Therefore, we need sum only over terms for which both $p_\mbf{k}^{(\tsf{b})} \neq 0$. We see that $Q(0)$ and $Q(1)$ do not equal $1$, since $|I_{\mbf{k}}|^2 < p_\mbf{k}^{(0)} p_\mbf{k}^{(1)}$ in general (see \cite{note2}). This is explained by the fact that the support of $\rho_0$, i.e., $\mathrm{span}\{ \psi_\mbf{k}^{(0)}\} \neq \mathrm{span}\{ \psi_{\mbf{k}}^{(1)}\}$, the support of $\rho_1$. Indeed, since  $\psi_{\mbf{k}}^{(0)} \not\propto \psi_{\mbf{k}}^{(1)}$ for any $\mbf{k}$, $\rho_0$ and $\rho_1$ do not have the same support on $\cl{H}_{\mbf{k}}$, and consequently also on $\oplus_{\mbf{k}} \cl{H}_\mbf{k}$. As a result, $Q(0) =  \mathrm{Tr}[\rho_{0}^0 \rho_{1}] =\mathrm{Tr} [P_0 \rho_1]\neq \mathrm{Tr}[I \rho_1] = 1$, for $P_0$ the projection operator onto the support of $\rho_0$ and $I$ the identity operator on $\oplus_{\mbf{k}} \cl{H}_\mbf{k}$. Similarly $\rho_1^0 = P_1 \neq I$ is the projector onto the support of $\rho_1$ and so $Q(1) = \mathrm{Tr}[\rho_0 P_1] \neq 1$.

For a  transmitted state $\ket{\Psi}= \otimes_{m=1}^M\ket{\psi_m}$ that is a product of $M$ signal-idler states $\ket{\psi_m}$, we have the multiplicative properties
\begin{align}
\cl{F}(\ket{\Psi}) = \prod_{m=1}^M \cl{F}(\ket{\psi_m})
\end{align}
and
\begin{align}
Q(s)(\ket{\Psi}) = \prod_{m=1}^M Q(s)(\ket{\psi}),
\end{align}
which simplify computations by converting the sum over vector $\mbf{k}$ to a product of scalar sums. We illustrate the results of this subsection in Section IV.D by applying them to quantum reading with the EPR state.

\section{Quantum Reading and Target Detection}

\subsection{Output State Fidelity and Related Error Probability Bounds} \noindent For the quantum reading and target detection scenarios, we have  $\Delta=\theta_1 - \theta_0 = 0$. In Section II, we obtained the lower bound \eqref{fidelitylowerbound} on the output state fidelity in terms of the input state's signal photon probability distribution $\{p_n\}$. For an NDS input state with that $\{p_n\}$, the fidelity \eqref{ndsfidelity} evaluates to
\begin{align}
\cl{F} =\left(\sum_{n=0}^{\infty} p_{n} (r_0r_1 + t_0 t_1)^{n}\right)^2,
\end{align}
which is \emph{exactly the general fidelity lower bound \eqref{Ofinal}-\eqref{fidelitylowerbound} of Section II} with $\Delta=0$. Thus, among all input states with a given $\{p_n\}$, the NDS states with that $\{p_n\}$ minimize the fidelity.

Further, since $\mu \equiv r_0r_1 + t_0 t_1 < 1$ when at least one $r_\tsf{b} <1$, we have by the convexity of the function $x \mapsto \mu^{x}$ and Jensen's inequality \cite{HLP88} that, for a given total signal energy $N_s$,
\begin{align} \label{minfidelity}
\cl{F} &= \left(\sum_{n=0}^{\infty} p_{n} (r_0r_1 + t_0 t_1)^{n}\right)^2 \\
&\geq\left( (r_0r_1 + t_0 t_1)^{\sum_{n=0}^{\infty} n p_{n}} \right)^2 \label{ineq}\\
&=(r_0r_1 + t_0 t_1)^{2N_s} \\
&\equiv \cl{F}_\tsf{min}(N_s). \label{Fmin}
\end{align}
The inequality \eqref{ineq} is an equality (at least at integer values of $N_s$) for precisely the states having $N_s$  total signal photons with probability one. The only NDS states with this property are the multimode Fock states $\ket{N_1}\otimes \cdots \otimes \ket{N_M}$ with total number of photons $ \sum_{m=1}^M N_m = N_s$. We have thus shown that a (possibly multimode) Fock state with $N_s$ photons minimizes the output fidelity at given $N_s$.

The above argument, employing convexity as it does, does \emph{not} go through when $\Delta \neq 0$. Moreover, even for $\Delta =0$, achieving minimum fidelity is \emph{not} equivalent to achieving minimum $\overline{P}_e$ as will be evident in the examples to follow in Sections IV.C-E.

\subsubsection{Error Probability Lower Bounds}
Since $\cl{F}_\tsf{min}(N_s)$ of \eqref{Fmin} is a lower bound on the output state fidelity for \emph{any} multimode pure input state $\ket{\psi}$ of energy $N_s$, we obtain the universal lower bound
\begin{align} \label{univlowerbound}
\frac{1}{2}\left(1- \sqrt{1-\cl{F}_\tsf{min}(N_s)}\right) &= \nonumber\\
\frac{1}{2}\left(1- \sqrt{1-(r_0r_1+t_0t_1)^{2N_s}}\right) &\leq \overline{P}_e[\ket{\psi}]
\end{align}
on the output error probability $\overline{P}_e[\ket{\psi}]$ using \eqref{fidelityboundsonhelstrom}.

For any input state $\ket{\Psi}$ consisting of $N$ copies of an $(M+M')$-mode signal-idler state $\ket{\psi}$
\begin{align}
\ket{\Psi}=\bigotimes_{n=1}^N \ket{\psi},
\end{align}
we may also obtain an upper bound on the \emph{quantum Chernoff exponent} $\xi_{\tsf{QCB}}= - \frac{\log Q[\Psi]}{N}$ \cite{Audenaert06}. It was shown in \cite{Audenaert06} that while the finite-$N$  behavior of $\overline{P}_e[\Psi]$  is complicated, $\overline{P}_e[\Psi]$ is asymptotically exponential in $N$ in the sense that
\begin{align}
 \lim_{N \rightarrow \infty} - \frac{\log \overline{P}_e[\Psi]}{N} = \xi_{\tsf{QCB}}.
\end{align}  Assuming $\ket{\psi}$ has finite average energy $N_s$ (so that $\ket{\Psi}$ has average energy $NN_s$), we have from the fidelity lower bound $\cl{F}_\tsf{min}(N_s) \leq \cl{F}[\psi]$ coupled with the lower bound of \eqref{fidelityboundsonhelstrom} that
\begin{align}
\frac{1}{4}\left[\cl{F}_\tsf{min}(N_s)\right]^N = \frac{1}{4}(r_0r_1 + t_0 t_1)^{2(N_s)\cdot N} \leq \overline{P}_e[\Psi],
\end{align}
from which the bound
\begin{align}
\xi_{\tsf{QCB}} \leq \log \left[(r_0r_1 + t_0 t_1)^{-2N_s}\right]
\end{align}
follows on taking logarithms.

\subsubsection{Reading of an Ideal Memory}
For the case of $R_1=1$ (called ``ideal memory'' in \cite{Pirandola11}), the channel $\cl{E}_1$ is simply the identity channel. Therefore, the return state $\rho_1$ is pure -- to wit, the transmitted state $\ket{\psi}\bra{\psi}$. In such a situation, the fidelity $\cl{F}[\ket{\psi}]$ equals the Chernoff bound $Q(\ket{\psi})$ \cite{Audenaert06}, so that we have the bound (which is stronger than the upper bound of \eqref{fidelityboundsonhelstrom}):-
\begin{align}
\overline{P}_e[\ket{\psi}] \leq \frac{1}{2} \cl{F}[\ket{\psi}].
\end{align}
In conjunction with the result of Section IV.A that the multimode Fock states attain the minimum fidelity, we have the remarkable consequence that among all the $2M$-mode signal-idler states with signal energy $N_s$, a signal mode Fock state with $N_s$ total photons has the best (lowest) Chernoff bound. See also the discussion in Section IV.C.1.

We mention in this connection that an optimization of the Chernoff bound was carried out in \cite{InvernizziParisPirandola10} fixing the total energy in signal and idler but restricting to the cases when the input state is a single-mode squeezed thermal state and an $M=1$ signal-idler two-mode squeezed thermal state -- it was shown that the single-mode and two-mode squeezed vacuum states minimize the ideal memory Chernoff bound in that class.

\subsubsection{Target Detection: No-Go Result for Large Loss}

The case of target detection corresponds to $r_0=0$ and $t_0=1$. Under these conditions, the universal lower bound \eqref{univlowerbound} reads
\begin{align} \label{detectionunivlowerbound}
\frac{1}{2}\left(1- \sqrt{1-(1-R_1)^{N_s}}\right) \leq \overline{P}_e[\ket{\psi}].
\end{align}
It is easy to check -- see \eqref{csmpe} -- that the error probability obtained from a pure coherent state of energy $N_s$ is
\begin{align} \label{detcspe}
\overline{P}_e[\mathrm{CS}] = \frac{1}{2}\Big[ 1- \sqrt{1-e^{-R_1 N_s}}\Big].
\end{align}
The case of large loss $R_1 \ll 1$ is of practical importance for standoff target detection. Because $e^{-R_1} \simeq 1 -R_1$ in \eqref{detcspe} under such conditions, we see that the error probability of a general input state of energy $N_s$, which is lower bounded by the LHS of \eqref{detectionunivlowerbound}, is not appreciably smaller than the coherent state error probability \eqref{detcspe}. We thus have a \emph{no-go result for appreciable quantum advantage in target detection under high loss conditions that applies to \emph{any} multimode input state}. Note that this does not contradict the $6$ dB advantage in the error exponent of the EPR state over coherent states claimed in \cite{Tan08} for high loss target detection because that analysis was carried out with the additional assumption of large thermal background noise in each signal mode.

The question arises if one can connect the fidelity with the Chernoff bound as we did in Section IV.A.2 above. If the input state has signal-idler entanglement, $R_0=0$ and $R_1 \neq 1$ imply that both $\rho_0$ and $\rho_1$ are mixed in general so that the argument connecting the fidelity and the Chernoff bound no longer applies. However, for the case of a pure transmitted state that is not entangled to any idler modes kept at the receiver -- this is called a Type I target detection scenario in \cite{yuen09} -- $\rho_0$ is a pure state, namely the vacuum state of the signal modes. Thus, it is again true that in a Type I target detection scenario, the number state transmitter yields the lowest Chernoff bound among all pure transmitted states of energy $N_s$. When $R_1 \ll 1$ prevails, the no-go result given above is in force even in this case. In such regimes, the coherent state performance is little different from the number state performance (see \eqref{fockdetpe} below). This holds even for moderately large $R_1$, as we will see in Section IV.E.1 (Refer Fig.~3).

In the remaining subsections, we compare in detail the performance of coherent states, number states, and the EPR states of \cite{Pirandola11}.

\subsection{Coherent States}\noindent
For a coherent state input of energy $N_s$, say the single-mode state (see \cite{note1}) $\ket{\sqrt{N_s}}$ of mean amplitude $\sqrt{N_s}$, the optimal error probability \eqref{helstrom} for discriminating $\rho_0$ and $\rho_1$ evaluates to
\begin{align} \label{csmpe}
\overline{P}_e[\mathrm{CS}] = \frac{1}{2}\Big[ 1- \sqrt{1-e^{-(r_1-r_0)^2N_s}}\Big].
\end{align}
The RHS is exactly the lower bound on the optimal error probability of \emph{any} classical state derived in \cite{Pirandola11} (Theorem 1 therein). Therefore, pure coherent states are the optimal classical states for quantum reading in the absence of added thermal noise. A physical realization of the optimal error probability (\ref{csmpe}) is provided by the so-called Dolinar receiver \cite{dolinar76} and some other receivers exist which approximate that performance. We discuss these issues briefly in Sec.~IV.E.2.

\subsection{Number States} \noindent Let us now consider the $\overline{P}_e$ obtained by transmitting number states. For a transmitted Fock state $\ket{\mbf{N}} = \ket{N_1}\otimes \cdots\otimes\ket{N_m}$ with $N_s = \sum_{m=1}^M N_m$, it is seen that  $\ket{\psi_{\mbf{k}}^{(\tsf{b})}} = \sqrt{p_\mbf{k}^{(\tsf{b})}} \ket{\mbf{N}-\mbf{k}},$ for $p_\mbf{k}^{(\tsf{b})}$ a product of binomial probabilities:-
\begin{align} \label{fockprob}
p_\mbf{k}^{(\tsf{b})} = \prod_{m=1}^M \left[\begin{pmatrix} N_m\\ k_m \end{pmatrix} R_{\tsf{b}}^{N_m -k_m} T_{\tsf{b}}^{k_m}\right].
\end{align}
The output states $\rho_0$ and $\rho_1$ commute, so that the techniques of Section III are not required to evaluate the performance. The optimal quantum measurement is photon counting on the individual modes followed by classical processing of the count results. It is easy to verify that the optimal decision rule is given by
\beq \label{fockdr}
 \frac {R_1^T \cdot T_1^{N_s -T}}{ R_0^T \cdot T_0^{N_s -T}} \begin{array}{c}
\mbox{\scriptsize say $\cl{E}_1$}\\ \geq \\ < \\
\mbox{\scriptsize say $\cl{E}_0$}\end{array} 1,
\eeq
for $T$ the observed total photon count (Recall that we are assuming $R_0 < R_1$). This is equivalent to the rule
\begin{align}
T \begin{array}{c}
\mbox{\scriptsize say $\cl{E}_1$}\\ \geq \\ < \\
\mbox{\scriptsize say $\cl{E}_0$}\end{array} T_*,
\end{align}
where the threshold $T_*$ equals
\begin{align}
T_* = N_s \frac{\ln\Big(\frac{T_0}{T_1}\Big)}{\ln\Big(\frac{R_1 T_0}{R_0 T_1}\Big)}.
\end{align}
Note that the decision rule is independent of the actual Fock state chosen as long as $N_s$ is fixed. Moreover, we can show from \eqref{fockprob} that the probability of counting $T$ photons is also independent of the distribution of input photons among the modes as long as the total number is $N_s$. Consequently, $\overline{P}_e$ is independent of the details of the distribution of the photons as well. This feature has practical implications that we discuss in Section IV.E.2. From \eqref{fockprob}-\eqref{fockdr}, $\overline{P}_e$ is easily computed numerically, as we will do in Section IV.E.1.

We also have the essentially classical Chernoff Bound
\begin{align} \label{numberstatechernoffbound}
\overline{P}_e \leq \frac{1}{2}\Bigg( R_0^{s_*}R_1^{1-s_*} +  T_0^{s_*}T_1^{1-s_*}\Bigg)^{N_s} \equiv \frac{1}{2} e^{-\xi_{\tsf{Num}} N_s},
\end{align}
where
\beq \label{focks*}
s_*= \Bigg(\ln \bigg[ \frac {-R_1 \ln(R_0/R_1)} {T_1 \ln(T_0/T_1)} \bigg]  \Bigg) \Bigg( \ln \bigg[  \frac{T_0R_1} {T_1R_0}\bigg]  \Bigg)^{-1}
\eeq
so that ${\xi_\tsf{Num}}$ is the number state Chernoff exponent. This bound is obtained by computing $Q(s)$ directly from the definition \eqref{chernoff} and finding the optimum exponent analytically. The above results apply when $R_0 \neq 0$ and $R_1 \neq 1$.

For a coherent state transmitter, the Chernoff bound is
\begin{align} \label{CSChernoff}
\overline{P}_e \leq \frac{1}{2} e^{-(r_1-r_0)^2N_s} \equiv \frac{1}{2} e^{-\xi_{\tsf{CS}} N_s},
\end{align}
for $\xi_{\tsf{CS}}$ the coherent state Chernoff exponent. A useful measure for quantifying the improvement obtainable from a number state transmitter from coherent state performance is the ratio of their Chernoff exponents, which we call the ``gain'' $G$:-
\begin{align} \label{gain}
G = \frac {\xi_{\tsf{Num}}}{\xi_{\tsf{CS}}}.
\end{align}
In Fig.~2, we plot $G$ against the reflectances $R_0$ and $R_1$, which shows that the gain is appreciably greater than unity only when both $R_0$ and $R_1$ are fairly large. Further performance comparisons at varying values of $N_s$ are made in Section IV.E.
\begin{figure}
\includegraphics[angle=90, trim=80mm 10mm 85mm 10mm, clip=true, width=0.5\textwidth]{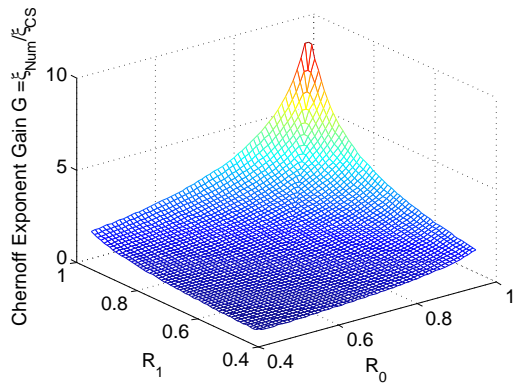}
\caption{(Color online) Number state vs. Coherent state Chernoff exponent Gain G of eq.~\eqref{gain} for values of $R_0$ and $R_1$ above $0.4$. Appreciable gain is seen for larger values of the reflectances.}
\end{figure}
\subsubsection{Ideal memory and Target Detection} \noindent For the case $R_0=0$ (target detection), we have the exact result
\begin{align} \label{fockdetpe}
\overline{P}_e = \frac{1}{2} T_1^{N_s}.
\end{align}
The optimum decision rule in this limit is to declare the target present if and only if the total count $T > 0$. In the opposite limit $R_1=1$ (the ideal memory of \cite{Pirandola11}), we likewise obtain
\begin{align} \label{fockidealpe}
\overline{P}_e = \ln \frac{1}{2} R_0^{N_s}
\end{align}
and the optimum decision rule is to declare $\tsf{b}=1$ if and only if $T = N_s$. The ideal memory case is exceptional in that, irrespective of the particular (pure) state transmitted, $\rho_1$ is a pure state -- the transmitted state itself. In such a situation, as mentioned in Section IV.A.2, the fidelity $\cl{F}$ equals the Chernoff bound \cite{Audenaert06}. Thus, the Chernoff bound reads
\begin{align}
\overline{P}_e \leq \frac{1}{2} \cl{F}
\end{align}
which, on comparing \eqref{Fmin} with $r_1=1$ to \eqref{fockidealpe}, is in fact an equality for Fock states.

\subsection{EPR State} \noindent
We now consider the $M$-mode EPR state with per-mode energy $N = N_s/M$ of \cite{Pirandola11}, which will afford an illustration of the techniques developed in Section III. The EPR state is an $M$-fold tensor product of a two-mode squeezed vacuum state which, for a single mode-pair, has the photon number representation
 \begin{align}
 \ket{\psi}_{\mathrm{EPR}}= \sqrt{\frac{1}{N+1}}\sum_{n=0}^{\infty} \left(\frac {N} {N+1}\right)^{n/2} \keti{n}\kets{n}.
\end{align}
$\ket{\psi}_{\mathrm{EPR}}$ is evidently an NDS state. For the case of no added thermal noise being considered in this paper, it was shown in \cite{Pirandola11} using the Bhattacharyya bound $Q(1/2)$ that, for $N_s> N_\mathrm{th}(R_0,R_1)$ -- a threshold total energy depending on the reflectances, there \emph{exists} an $M$ for which the classical error probability is worse than the EPR Bhattacharyya bound (Theorem 2 of \cite{Pirandola11}). For the case of ideal memory, a similar threshold theorem was shown to hold with $N_\mathrm{th} = 1/2$ for \emph{all} $M$ greater than a threshold $\overline{M}$ (Theorem 3 of \cite{Pirandola11}). We show here that the techniques of Section III may be used to derive these results in an alternative manner, as well as providing numerical results at chosen $N$ and $M$.

We first obtain the output fidelity and $Q(s)$ for the state $\ket{\psi}_\mathrm{EPR}$, i.e., for  $M=1$.  The photon probability distribution $p_k^{\tsf{b}}$ of \eqref{pkb} in the output environment mode is in the Bose-Einstein form of a thermal state with $T_\tsf{b}N$ average photons:-
\begin{align}
p_{k}^{(\tsf{b})} = \frac {1}{T_\tsf{b} N +1} \left( \frac{T_\tsf{b} N}{T_\tsf{b} N +1}\right)^k.
\end{align} Evaluating \eqref{Ik} yields
\begin{align}
I_k = \frac {(t_0 t_1 N)^k} {[(1-r_0r_1)N +1]^{k+1}}.
\end{align}
We may now use \eqref{ndsfidelity} and \eqref{ndschernoffquantity} to get:-
\begin{align} \label{eprfidelity}
\cl{F}|_{M=1}= \frac {1} {[(1-r_0r_1 - t_0 t_1)N +1]^2}
\end{align}
and
\begin{align} \label{eprChernoff}
Q(s)|_{M=1} = \frac{1}{C\alpha^{s} - D \beta^s},
\end{align}
with
\begin{align}
\alpha &= \frac{T_0N +1}{T_1N+1} > 1, \label{alpha} \\
\beta & = \frac{T_0}{T_1} > 1, \label{beta}\\
C &=\frac{[(1-r_0r_1)N+1]^2}{T_0N+1}, \label{C}
\end{align}
and
\begin{align}
D=T_1N. \label{D}
\end{align}
We may verify that the inequalities
\begin{align} \label{ineq1}
\alpha < \beta
\end{align}
and
\begin{align} \label{ineq2}
C>D
\end{align}
hold \cite{note3}. For a given $N$, obtaining the Chernoff bound entails finding the exponent $s_*$ that minimizes \eqref{eprChernoff}. We know that $Q(s)$ is a convex (and continuous \cite{note4}) function of $s$ in $[0,1]$ \cite{Audenaert06}. It is, from \eqref{eprChernoff}, evidently also twice differentiable in $s$. Accordingly, two cases logically arise depending on the sign of $Q'(0)$:-
\begin{enumerate}
\item If $Q'(0)\geq 0$, $Q(s)$ is an increasing function of $s$, so that $s_*=0$.
\item If $Q'(0) < 0$, then $s_*$ may be found by setting $Q'(s_*)=0$. If  this equation has no solution in [0,1], $s_*=1$.
\end{enumerate}

Differentiating \eqref{eprChernoff}, we find that
\begin{align}
Q'(s) = (C\alpha^s-D\beta^s)^{-2} [ (\ln \beta^D)\beta^s -(\ln \alpha^C) \alpha^s].
\end{align}
Consequently, the condition for deciding among the above two cases is
\beq \label{EPRCases}
\delta \equiv \frac{\beta^D} {\alpha^C}  \begin{array}{c}
\mbox{\scriptsize Case $1$}\\ \geq \\ < \\
\mbox{\scriptsize Case $2$}\end{array} 1.
\eeq

In the event of Case $2$, setting $Q'(s_*)=0$ gives $s_*$ as
\begin{align} \label{eprcase2s*}
s_* = \frac {\ln \left[ \frac{\ln\alpha^{C}} {\ln \beta^D} \right]} {\ln \left[ \frac{\beta}{\alpha}\right]}.
\end{align}
In the event that the RHS of \eqref{eprcase2s*} is greater than $1$, we have $s_*=1$.

Finally, from the multiplicative properties of the fidelity and the Chernoff-type quantities, we have that, for the $M$-mode input state $\ket{\psi}= \otimes_{m=1}^{M}\ket{\psi}_{\mathrm{EPR}}$, the fidelity equals
\begin{align} \label{EPRfidelity}
\cl{F}[\mathrm{EPR}] = \left[(1-r_0r_1-t_0t_1)N +1\right]^{-2M},
\end{align}
and the Chernoff bound is
\begin{align} \label{EPRChernoff}
\overline{P}_e \leq \frac{1}{2}Q[\mathrm{EPR}]=\frac{1}{2}Q(s_*) = \frac{1}{2} [C\alpha^{s_*}-D\beta^{s_*}]^{-M},
\end{align}
for the $s_*$ obtained from the above case analysis.

When the input state $\ket{\psi}$ is a multimode \emph{Gaussian state}, i.e., a state whose Wigner function is a Gaussian probability distribution (see e.g.,\cite{BraunvLoock05}), so are $\rho_0$ and $\rho_1$ since $\cl{E}_\tsf{b}^{\otimes^M} \otimes \tsf{id}^{\otimes^{M'}}$ is a linear, and hence Gaussian, channel. The input state $\ket{\psi}_\mathrm{EPR}$ is a Gaussian state. We could therefore use the general Gaussian state technique of \cite{ParaScuta00} to derive the fidelity between $\rho_0$ and $\rho_1$. Similarly, the technique of symplectic diagonalization \cite{Pirandola08} may be used to derive $Q(s)$, as it was in \cite{Pirandola11}.

Let us now connect these results to Theorems 2 and 3 of \cite{Pirandola11} -- the statements of these theorems were reviewed in the first paragraph of this subsection.  Theorem 2 makes essential use of the large $M$ limit of the Bhattacharyya bound $Q(1/2)$ given by \eqref{chernoff}. Some lengthy but straightforward algebra verifies that the $M \rightarrow \infty$ limit (at constant $N_s = MN$) of the $M$-pair Bhattacharyya bound $B(N_s)$
\begin{align} \label{Bhatta}
B(N_s) = \frac{1}{2} \lim_{M \rightarrow \infty} \left[ Q(1/2)|_{M=1}\right]^M
\end{align}
for $ Q(1/2)|_{M=1}$ obtained from \eqref{eprChernoff} is identical to the corresponding result (117) of \cite{Pirandola11supp} after adjusting for differing notation (see \cite{note5}). The same reasoning in \cite{Pirandola11supp} that establishes the ``Threshold energy'' theorem (Theorem 2 of \cite{Pirandola11}) may then be carried out from \eqref{Bhatta} to give an identical threshold energy as before.

Consider now the case of an ideal memory with $T_1=0$ for comparison to Theorem 3 of \cite{Pirandola11}. That theorem depends on the expression (129) of \cite{Pirandola11supp} for the Chernoff bound which we rederive using the method of this section. The same analysis as in \cite{Pirandola11supp} can then be used to reproduce the result of Theorem 3. From \eqref{D}, we have $D=0$ so that \eqref{eprChernoff} is clearly minimized at $s_*=1$ so that the $M$-mode Chernoff bound is
\begin{align} \label{idealmemChernoffbd}
\overline{P}_e \leq \frac{1}{2}Q[\mathrm{ideal}] = \frac{1}{2}[(1-r_0)N+1]^{-2M}
\end{align}
which agrees with (129) of \cite{Pirandola11supp}-\cite{note5}. Further, we see that the fidelity \eqref{EPRfidelity} also equals $Q[\mathrm{ideal}]$ as it should because $\rho_1$ is a pure state. In the general case of $R_1 \neq 1$, we can study the behavior of the Chernoff bound as a function of $M$ and $N_s$ by numerically obtaining $s_*$ for each value of these parameters.

\subsection{Comparison of Coherent State, Number State, and EPR State Transmitters}

\subsubsection{Numerical Comparison of Error Probability}
\noindent
In this subsection, we compare quantitatively the error probability performance  of the three types of states considered in the previous subsections for target detection, and reading of non-ideal and ideal memories. The representative plots below show the error probability on the y-axis in logarithmic scale against the total average signal energy $N_s$ on the x-axis. We assume the number of modes $M=50$. We reiterate that the number of modes has no effect on either the coherent state or number state performance, which depend on $N_s$ alone. For the EPR state, varying $M$ changes the performance (as given by the Chernoff bound), although the change is not appreciable once $M$ is around $20-30$. Thus, the plots given here are fairly representative of the best possible EPR state performance.

We summarize how the plots were made. For each $N_s$, the universal lower bound of \eqref{univlowerbound} was plotted in Figs.~2-5.  The coherent state error probability is given by the closed form expression \eqref{csmpe}. The number state error probability was calculated numerically for each value of $N_s$ using the count probability distribution \eqref{fockprob} and the decision rule \eqref{fockdr} for the cases of reading of non-ideal memories. The number state Chernoff bound \eqref{numberstatechernoffbound}-\eqref{focks*} was also plotted for these cases. Closed-form expressions \eqref{fockdetpe} and \eqref{fockidealpe} were used for the detection and ideal memory cases. The EPR state fidelity expression \eqref{EPRfidelity} in conjunction with \eqref{fidelityboundsonhelstrom} gives the EPR fidelity lower bound
\begin{align}\label{EPRLB}
\frac{1}{2}(1- \sqrt{1-\cl{F}[\mathrm{EPR}]})\leq \overline{P}_e[\mathrm{EPR}].
\end{align}
Finally, for each value of $N_S$, the EPR Chernoff bound \eqref{EPRChernoff} was calculated for the detection and non-ideal memory cases using the procedure for obtaining $s_*$ described in Sec.~IV.D.
\begin{figure}
\includegraphics[trim= 20mm 79mm 25mm 80mm, clip=true, width = 0.5\textwidth]{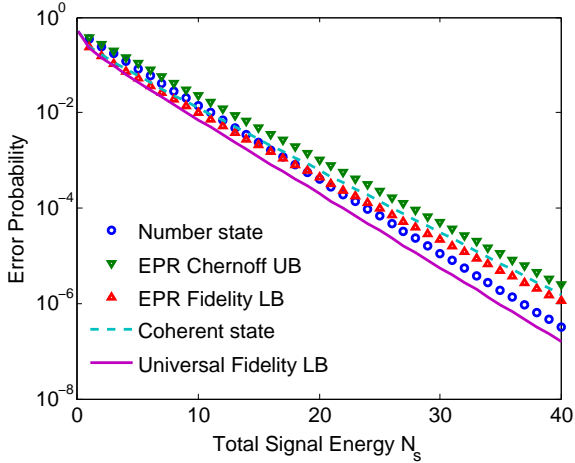}
\caption{(Color online) Error Probability Bounds versus $N_s$ for target detection with $R_0=0$ and $R_1=0.3$. The number of modes $M = 50$ for the EPR state curves.}
\end{figure}

\begin{figure}
\includegraphics[trim= 20mm 70mm 25mm 75mm, clip=true, width = 0.5\textwidth]{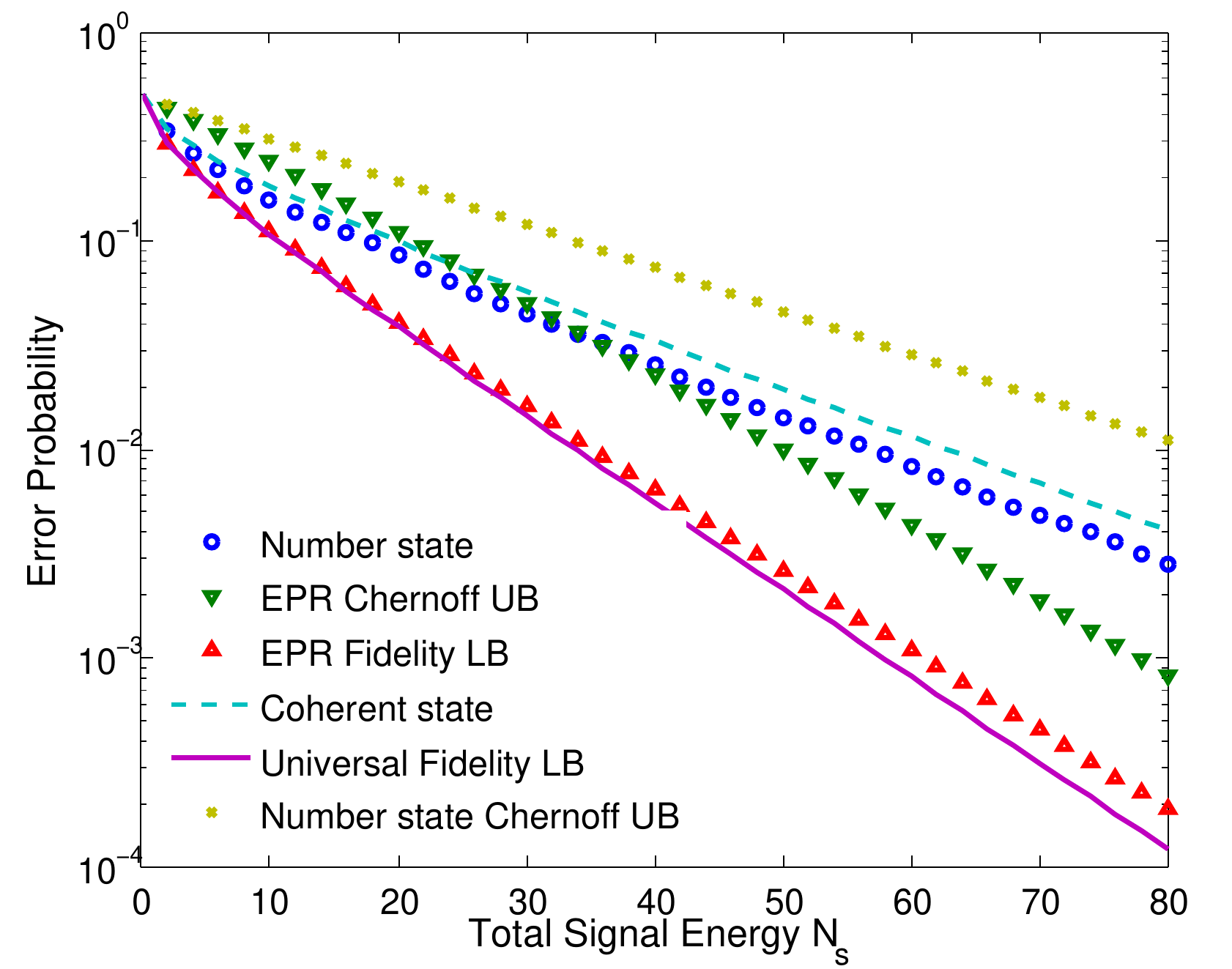}
\caption{(Color online) Error Probability Bounds versus $N_s$ for quantum reading with $R_0=0.3$ and $R_1=0.6$. The number of modes $M = 50$ for the EPR state curves.}
\end{figure}

\begin{figure}
\includegraphics[trim= 20mm 70mm 25mm 75mm, clip=true, width = 0.5\textwidth]{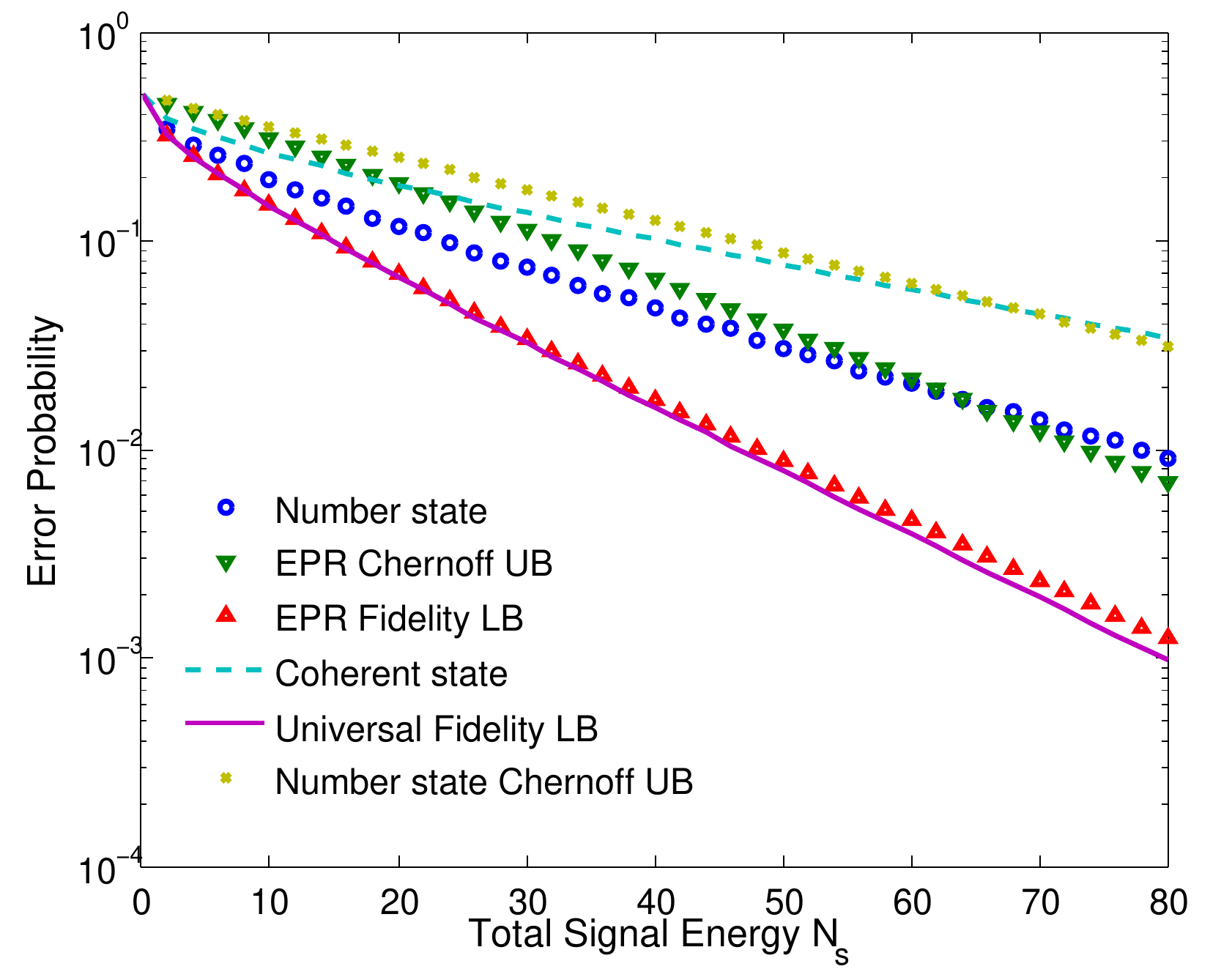}
\caption{(Color online) Error Probability Bounds versus $N_s$ for quantum reading with $R_0=0.5$ and $R_1=0.75$. The number of modes $M = 50$ for the EPR state curves.}
\end{figure}

\begin{figure}
\includegraphics[trim= 28mm 77mm 29mm 83mm, clip=true, width = 0.5\textwidth]{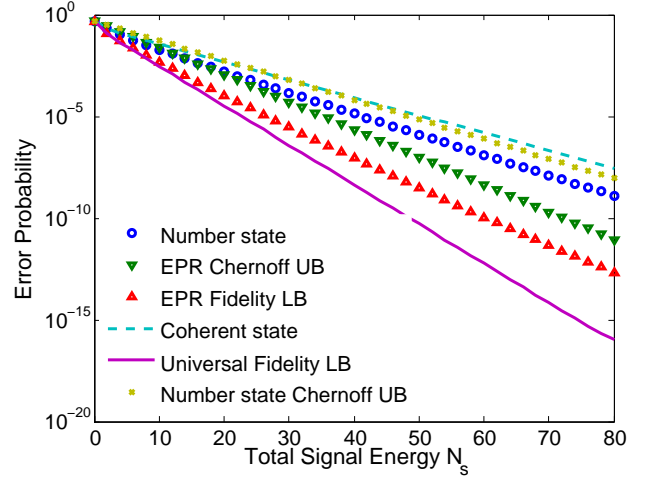}
\caption{(Color online) Error Probability Bounds versus $N_s$ for quantum reading with $R_0=0.2$ and $R_1=0.8$. The number of modes $M = 50$ for the EPR state curves.}
\end{figure}

In all the figures, we see that the number state and EPR state transmitters eventually outperform the coherent state transmitter, in tune with the conclusions of \cite{Pirandola11}. In the target detection case (Fig.~3), the number state also outperforms the EPR state lower bound, although the performance difference between the three states is not appreciable. We see that, even for an $R_1$ that is much larger than that expected in a realistic target detection scenario, the coherent state performance is not appreciably worse than the number state performance. As $R_0$ increases, the perfomance gain over classical increases also as evinced in Figs.~4-7. Fig.~4 shows a case where the difference $R_1 -R_0$ is small and the reflectances themselves are not very high. In such cases, the number state and coherent state performances are not appreciably different. However, the EPR Chernoff bound drops below the number state performance for $N_s$ greater than about $40$ photons. Fig.~5 also shows a case of small $R_1 -R_0$, but the reflectances themselves are appreciable. We see that the nonclassical transmitters' gain over the coherent state increases. Further, the crossover between the number state performance and the EPR Chernoff bound occurs later (at about $65$ photons) and the slopes of these two curves are less different than in Fig.~4. Fig.~6 represents distinguishing channels with large $R_1 -R_0$. We see that the nonclassical gain over the coherent state transmitter is even greater, with the EPR state doing better than the number state -- the crossover of the number state performance with the EPR Chernoff bound now occurs at about $20$ photons. Finally, Fig.~7 represents the reading of an ideal memory with $R_0=0.5$. The nonclassical gain is now very large, and the number state transmitter performs better than the EPR state, as evinced by the fact that it lies below the EPR state lower bound. All the plots are consistent with the number state vs. coherent state Chernoff exponent gain of Fig.~2 and Fig.~7 confirms and strengthens the conclusion of Section IV.A.2 that the number state transmitter has the lowest Chernoff bound for reading of an ideal memory.
\begin{figure}
\includegraphics[trim= 20mm 79mm 25mm 80mm, clip=true, width = 0.5\textwidth]{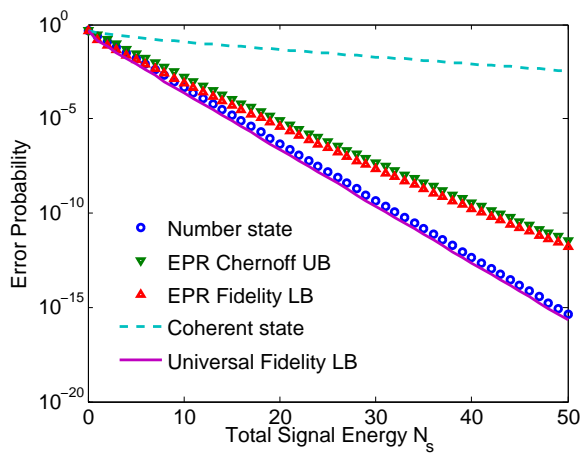}
\caption{(Color online) Error Probability Bounds versus $N_s$ for reading of an ideal memory with $R_0=0.5$ and $R_1=1$. The number of modes $M = 50$ for the EPR state curves.}
\end{figure}
\subsubsection{Experimental Considerations}
\noindent
In this section, we indicate some technological considerations regarding the availability of sources and detectors that bear upon the implementation of quantum reading with coherent, number, and EPR states.

While coherent sources of any energy are readily available, the optimal Helstrom detector of \cite{dolinar76} is not easy to realize as it employs feedback in addition to linear optics and photodetection. Nevertheless, the Kennedy receiver \cite{Kennedy73} and the so-called Optimum Displacement Receiver (ODR) \cite{Takeoka08} achieve the same error exponent and are more easily implemented since they do not involve feedback. Indeed, the ODR was recently demonstrated \cite{Tsujino11} with an overall detection efficiency of $\sim 90 \%$.

A single-mode number state with $N_s >1$ is hard to generate with current technology. However, we saw in Section~IV.C that a state consisting of $N_s$ spatial or temporal modes each of which is in a single photon state, i.e., a state of the form  $\ket{1}\otimes \cdots\otimes \ket{1}$ with $N_s$ total photons, has identical performance. As a large variety of single photon sources are available (see, e.g., \cite{LounisOrrit05}), the generation of multimode Fock state with many modes each in a single photon state does not appear problematic with current technology. The optimal Helstrom measurement is photon counting on the individual modes followed by classical processing. While high quantum efficiency photon number-resolving detectors are still a developing technology (see, e.g., \cite{Rosenberg05}) and require cooling to superconducting temperatures, the source described above consisting of multiple single-photon states requires only standard single-photon avalanche diode (SPAD) \cite{SalehTeich07} technology with either one or many detectors depending on the mode implementation. Detector quantum efficiencies and other system losses act as multiplicative factors to $R_0$ and $R_1$ and do not essentially change the analysis described in Section IV.C, as also is the case for the coherent state transmitter result of Section IV.B. SPADs, for instance, can attain quantum efficiencies of around $\eta \sim 0.75$. After adjusting $R_0$ and $R_1$ to account for system inefficiencies, it is fair to say that the optimum Helstrom receiver for quantum reading with number states is realizable with current technology.

Generating multimode EPR states of high total energy $N_s$ and large $M$ also does not seem to present a huge experimental problem (see, e.g., \cite{WongShapiroKim06}). Unfortunately, the optimum Helstrom detector, while given abstractly by \eqref{Pi0}-\eqref{Pi1}, is an entangling measurement over the $M$ modes and has no known concrete realization. A realizable suboptimal measurement involving homodyne detection was proposed in \cite{Pirandola11,Pirandola11supp} and was shown, significantly, to also outperform the coherent state transmitter. However, it appears that this comparison was made in \cite{Pirandola11supp} only for rather high $R_0$ and $R_1$. The effects of non-unity homodyne detector quantum efficiency, analogously to the Fock state and coherent state cases above, have also not yet been considered in that measurement. Moreover, given the comparisons made in this paper, it is of interest to see how the suboptimal measurement compares to the number state performance for reasonable values of $N_s$ and $M$.

\section{Conclusion} \noindent
The problem of distinguishing two optical beam splitter channels using multimode signal-idler entangled pure states was considered. A general lower bound on the output state fidelity and minimum error probability for any such input was derived. For Number-Diagonal-Signal States, series formulae for the optimum error probability, the output state fidelity, and the Chernoff-type upper bounds were derived. For quantum reading and target detection, for a given signal photon probability mass function, the fidelity bound was shown to be attained by NDS states, with multimode Fock states minimizing the bound for a given total photon number. For reading of an ideal memory with arbitrary states and for Type I (signal-only) target detection, the number state was shown to yield the best Chernoff bound among all states of given energy. For target detection under high loss conditions, a general no-go result for quantum advantage over coherent states was obtained. The above results were applied to quantitatively studying the performance gains over classical states obtainable by number state and EPR state transmitters, which were found to outperform the classical transmitters to varying degrees over a wide range of reflectances. The experimental outlook on realizing the optimal measurement for the number state transmitter was argued to be good. It is of interest to compare the performance, taking into account realistic experimental parameters, of the non-Helstrom measurement on the EPR state transmitter suggested in \cite{Pirandola11} and the number state transmitter suggested here. Finally, the techniques developed here are likely to prove useful for other interesting problems fitting the same framework, e.g., the lossy discrimination of optical phase shift channels.

\section{Acknowledgements} \noindent
The author is grateful to Alessandro Bisio, Michele Dall'Arno, Giacomo M.~D'Ariano, Saikat Guha, Bhaskar Mookerji, Stefano Pirandola, Jeffrey H.~Shapiro, Franco N. C.~Wong, Brent J.~Yen, and Horace P.~Yuen for useful discussions. This material is based upon work funded by DARPA's Quantum Sensor Program, under AFRL Contract No. FA8750-09-C-0194.


\begin{thebibliography}{10}
\bibitem{note1} It is enough to consider this essentially single-mode coherent state since its output under $\cl{E}_\tsf{b}^{\otimes ^M}$ is readily converted through an $M$-mode beam splitter transformation to the output under the same channel of any other product coherent state of the same total energy.
\bibitem{noteondef} We note that the definition \eqref{fidelity} of fidelity agrees with that in \cite{Uhlmann76,Jozsa94} but is the square of the definition in \cite{nielsen00}.
\bibitem{caveat} While the addition of effectively thermal noise from decoherence effects and by stray photons of the order of $10^{-5}$ photons per mode was also considered in \cite{Pirandola11}, we do not do so here because it appears difficult to carry over the analysis presented in this paper to the more general case of added thermal noise.
\bibitem{note2} \eqref{selfip} and \eqref{mutualip} together imply by the Cauchy-Schwarz inequality that $|I_\mbf{k}| \leq \sqrt {p_\mbf{k}^{(0)}p_\mbf{k}^{(1)}}$ so that the expression under the square root in \eqref{Deltarhok} and \eqref{Pebarnds} is non-negative.
\bibitem{note3} From \eqref{C}-\eqref{D}, $C>D$ can be shown to be equivalent to a quadratic polynomial in $N$ with coefficients which are functions of the beam splitter parameters being positive. Using $R_0 <R_1$ along with the arithmetic mean-geometric mean inequality, each of the coefficients may be shown to be positive, yielding the result.
\bibitem{note4} Indeed, the \emph{smallest} value taken by the denominator of \eqref{eprChernoff} for $s \in [0,1]$ is lower bounded by $C-D\beta$ which can be shown to be positive. Hence, $Q(s)$ remains bounded in $[0,1]$.
\bibitem{note5} In keeping with usual optics notation, we have used $R_\tsf{b}$ in lieu of $r_\tsf{b}$ of \cite{Pirandola11} for the beam splitter (energy) reflectances, reserving the lower case symbols for the magnitudes of field reflection coefficients. We use $N_s$ for the average total signal energy throughout this paper which corresponds to $N$ of \cite{Pirandola11}. We use $N$ in Section IV.D for the per-mode EPR state energy so that our $N=N_s/M$.
\end{thebibliography}
\end{document}